\documentclass[aps,twocolumn,showpacs,longbibliography,nofootinbib,groupedaddress]{revtex4-2}
\usepackage{CJK} 
\usepackage{graphicx}
\usepackage{bm}
\usepackage{multirow}
\usepackage{array}
\usepackage{graphicx}
\usepackage{amsmath}
\usepackage{txfonts}
\usepackage{mathrsfs}
\usepackage{subfigure}
\usepackage{amssymb}
\usepackage{caption}
\usepackage{float}
\usepackage{booktabs}
\usepackage{diagbox}
\usepackage{setspace}
\usepackage{enumerate}
\usepackage{color}
\usepackage{tcolorbox}
\usepackage{inconsolata}
\usepackage{orcidlink}
\definecolor{mygreen}{cmyk}{1,0.02,0.,0.42}        
\definecolor{myblue}{cmyk}{0.38,0.18,0.,0.46}

\usepackage{slashed}
\usepackage{lipsum}
\usepackage{hyperref}
\usepackage{url}
\hypersetup{hypertex=true,
	colorlinks=true,
	linkcolor=mygreen,
	anchorcolor=red,
	citecolor=blue}

\begin{document}
	\setlength{\textfloatsep}{5pt}
	\begin{CJK*}{UTF8}{gbsn}
	\title{$U_A(1)$ symmetry restoration at high baryon density}
	\author{Jianing Li ({\CJKfamily{gbsn}李嘉宁})\orcidlink{0000-0001-7193-7237}}
	\email{ljn18@tsinghua.org.cn}
	\affiliation{Department of Physics, Tsinghua University, Beijing 100084, China}
	\author{Jin Gui ({\CJKfamily{gbsn}桂进)}}
	\affiliation{Department of Physics, Tsinghua University, Beijing 100084, China}
	\author{Pengfei Zhuang ({\CJKfamily{gbsn}庄鹏飞})\orcidlink{0000-0002-9639-1493}}
	\affiliation{Department of Physics, Tsinghua University, Beijing 100084, China}
	\date{\today}
	\begin{abstract}
		We study the relation between chiral and $U_A(1)$ symmetries in the quark-meson model. Although quarks and mesons are described in mean field approximation, the topological susceptibility characterizing the $U_A(1)$ breaking comprises two components: one controlled by the condensate and the other by the meson fluctuation. The $U_A(1)$ restoration is governed by the competition of these components. In a hot medium, the condensates melt. However, the fluctuation is enhanced. Therefore, the $U_A(1)$ symmetry cannot be solely restored via the temperature effect. Nevertheless, the baryon density reduces the condensates and fluctuation, and thereby, the $U_A(1)$ symmetry can only be restored in a dense or dense and hot medium. The strange condensate plays a weak role in the susceptibility, and the chiral and $U_A(1)$ symmetry restorations occur almost at the same critical point.
	\end{abstract}
	\keywords{$U_A\left(1\right)$ symmetry, chiral phase transition, finite temperature field theory}
	\maketitle
\end{CJK*}
	
\section{Introduction}
\label{s1}
In quantum chromodynamics (QCD) which refers to the theory for strong interaction, the chiral symmetry is broken at the (classical) mean field level~\cite{Cheng:2000ct}, and the $U_A(1)$ symmetry is broken at the (quantum) loop level due to the nontrivial topology of the principle bundle of the gauge field~\cite{tHooft:1976rip,tHooft:1976snw,Leutwyler:1992yt}. It is widely accepted that a strongly interacting system should be in a symmetric state when the temperature of the system is sufficiently high. Hence, the chiral symmetry~\cite{Pisarski:1983ms} and $U_A(1)$ symmetry~\cite{Schafer:1996hv} are expected to be restored in a hot medium. However, based on lattice simulations, while the chiral symmetry is smoothly restored at the critical temperature $T_c\sim 155$ MeV~\cite{Aoki:2006we}, the $U_A(1)$ symmetry is only partially restored by the temperature effect but still broken at temperatures above $T_c$, even in the chiral limit~\cite{Cossu:2013uua,Ding:2020xlj,Kaczmarek:2021ser,Dentinger:2021khg}. Many model calculations~\cite{Fukushima:2001hr,Fukushima:2001ut,Costa:2002gk,Costa:2003uu,Costa:2004db,Costa:2005cz,Chen:2009gv,Brauner:2009gu,Contrera:2009hk,Ruivo:2011fg,Jiang:2012wm,Ruivo:2012xt,Xia:2013caa,Jiang:2015xqz,Rai:2018ufz,Li:2019chs,Kawaguchi:2020kdl,Kawaguchi:2020qvg,Cui:2021bqf,Fejos:2021yod} at finite temperature with 2 or 2+1 flavors and experimental measurements in high energy nuclear collisions~\cite{Csorgo:2009pa} support the lattice results. Hence, a natural question is then raised: can the $U_A(1)$ symmetry be restored? If yes, what is the condition?

Unlike the temperature effect that gradually alters chiral symmetry, baryon density results in a first-order chiral phase transition, both in the chiral limit and in the real world~\cite{Ejiri:2008xt}. The density effect for a fermion system is a pure quantum effect induced by the Pauli exclusion principle~\cite{Fukushima:2010bq}. The abrupt shift of the chiral condensate from a nonzero value to zero in the chiral limit or from a higher to a lower value in the real world is driven by the system's pronounced Fermi surface. We expect that this jump can aid in restoring the $U_A(1)$ symmetry at high baryon density. The insights on $U_A(1)$ breaking at finite baryon density are relatively rare. Considering the nuclear collisions on plan which can create high baryon density~\cite{An:2021wof}, relevant study on the change in $U_A(1)$ symmetry at finite baryon chemical potential $\mu_B$ is required. The goal of this study is to examine the relation between chiral symmetry and $U_A(1)$ symmetry in a hot and dense medium.

Blocked by the sign problem, lattice QCD simulation loses its efficacy at large $\mu_B$~\cite{Barbour:1997ej}, and we have to consider an effective model to account for the non-perturbative calculations. There are two types of models that effectively describe the chiral and $U_A(1)$ symmetries. One approach operates at the quark level, as exemplified by the Nambu--Jona-Lasinio (NJL) model~\cite{Nambu:1961tp,Nambu:1961fr}, while the other functions at the hadron level, akin to the quark-meson model~\cite{levy1967currents,Lenaghan:2000ey,Schaefer:2008hk}. In the NJL model, quarks are elementary particles, and hadrons are treated as quantum fluctuations above the mean field via random phase approximation~\cite{Klevansky:1992qe}. In the quark-meson model, quarks and hadrons are elementary degrees of freedom, which largely simplify the derivation of mesonic correlation functions in the calculation of topological susceptibility for the study of $U_A(1)$ symmetry.

The paper is organized as follows. In Sec. \ref{s2}, we briefly review the (2+1)-flavor quark-meson model, derive the topological susceptibility $\chi$, which is the order parameter for the phase transition from $U_A(1)$ symmetry breaking to its restoration, and diagrammatically analyze the condition for the $U_A(1)$ restoration in the quark-meson and NJL models. In Sec. \ref{s3}, we analytically and numerically calculate the susceptibility and mass splitting between $\eta$ and $\eta'$ mesons at finite temperature and baryon density, wherein the latter is often used to measure the degree of $U_A(1)$ breaking. Finally, we summarize the paper in Sec. \ref{s4}.
\section{Topological susceptibility}
\label{s2}
The topological susceptibility $\chi$ is the order parameter of a quantum phase transition. Based on the QCD Lagrangian density with a $\theta$-term~\cite{Callan:1976je,Jackiw:1976pf}
\begin{equation}
\mathcal {L} = -{1\over 4} F_{\mu\nu}^a F_a^{\mu\nu}+\bar\psi\left(\mathrm{i}\gamma^\mu D_\mu-m\right)\psi+\theta Q.
\end{equation}
With the gluon field tensor $F^a_{\mu\nu}$ in both Dirac and color spaces ($\mu,\nu=0,1,2,3,\ a=0,1,\cdots,8$), we consider the covariant derivative $D_\mu$. The quark mass matrix, denoted as $m=\text{diag}(m_q,m_q,m_s)$, is defined in flavor space with light quarks $q=u, d$ and strange quark $s$. We also account for vacuum angle $\theta$ and topological charge density $Q$, which can be defined as
\begin{equation}
Q(x) = \frac{g^2}{32\pi^2}F^a_{\mu\nu}(x)\tilde F_a^{\mu\nu}(x),
\end{equation}
the vacuum energy density of QCD is the path integral of the action of the system,
\begin{equation}
\varepsilon=-\frac{1}{V}\ln\int\mathcal{D}A_\mu\mathcal{D}\bar\psi\mathcal{D}\psi \mathrm{e}^{\int \mathrm{d}^4x\mathcal{L}}
\end{equation}
in four dimensional space volume $V$, and the susceptibility $\chi$ can be formally defined as
\begin{eqnarray}
\label{chi}
\chi = {\partial^2\varepsilon\over\partial\theta^2}\Big|_{\theta=0}
= \int \mathrm{d}^4x\langle\mathcal{T}\left[Q(x)Q(0)\right]\rangle_\mathrm{connected},
\end{eqnarray}
where $\mathcal{T}$ denotes the time-ordering operator, $\langle\cdots\rangle$ denotes ensemble average, and only connected diagrams contribute to the susceptibility.

The topological charge $Q$ corresponds to an infinite small $U_A(1)$ transformation for the quark field, $\psi\to \mathrm{e}^{\mathrm{i}\theta\gamma_5T_0}\psi\to \psi-\mathrm{i}\theta\gamma_5\psi/\sqrt 6$, where $T_a$ denotes the Gell-Mann matrices with the unit matrix $T_0=\sqrt{1/6}$, normalization $\text{Tr}(T_aT_b)=\delta_{ab}/2$, equations $\{T_a,T_b\}=d_{abc}T^c$ and $[T_a,T_b]=\mathrm{i}f_{abc}T^c$, and symmetric and anti-symmetric structure constants $d_{abc}$ and $f_{abc}$ ($d_{ab0}=\sqrt{2/3}\delta_{ab}$ and $f_{ab0}=0$) . Under this transformation, the axial current $J^5_\mu=\bar\psi\gamma_\mu\gamma_5\psi$ is not conserved:
\begin{equation}
\label{j51}
\partial^\mu J^5_\mu=2N_fQ+2\mathrm{i}\bar\psi m\gamma_5\psi.
\end{equation}

We now derive the hadronic version of the susceptibility (\ref{chi}) in the $SU_3\times SU_3$ quark-meson model, following Ref.\cite{Jiang:2015xqz}. The model is defined as~\cite{levy1967currents,Lenaghan:2000ey,Schaefer:2008hk}
\begin{equation}
	\mathcal{L}_{QM} = \mathcal{L}_Q+\mathcal{L}_M
\end{equation}
with the meson section
\begin{eqnarray}
	\label{lm}
	\mathcal{L}_M &=& \text{Tr}\left(\partial_\mu\phi^\dagger\partial^\mu\phi\right)-\lambda^2\text{Tr}\left(\phi^\dagger\phi\right)-\lambda_1\left(\text{Tr}(\phi^\dagger\phi)\right)^2\\
	&& -\lambda_2\text{Tr}\left(\phi^\dagger\phi\right)^2+\text{Tr}\left(H(\phi^\dagger+\phi)\right)+c\left(\text{det}(\phi^\dagger)+\text{det}(\phi)\right)\nonumber
\end{eqnarray}
and quark section
\begin{equation}
\label{lq}
\mathcal{L}_Q = \bar\psi\left[\mathrm{i}\gamma^\mu\left(\partial_\mu-\mathrm{i}{\mu_B\over 3}\delta_\mu^0\right)-gT^a\left(\sigma_a+\mathrm{i}\gamma_5\pi_a\right)\right]\psi.
\end{equation}

In the meson part, $\phi$ denotes a complex $3\times 3$ matrix composed of scalar and pseudoscalar nonets $\sigma_a$ and $\pi_a$, $\phi=T^a\phi_a=T^a(\sigma_a+\mathrm{i}\pi_a)$, $\lambda^2$ is the mass parameter, and the coupling constants $\lambda_1$ and $\lambda_2$ characterize the interaction among the mesons. Given that we do not have strict chiral symmetry in the real world, the explicit symmetry breaking enters the model by introducing two external sources $h_0$ and $h_8$ via $H=\text{diag}(h_0, h_0, h_8)$. We are concerned with $U_A(1)$ symmetry, which is explicitly broken by the determinant term with an anomaly parameter $c$.

In the quark part, $\psi$ is the quark field with three flavors $N_f=3$ and three colors $N_c=3$, $\mu_B$ $(\mu_B/3)$ denotes the baryon (quark) chemical potential, and $g$ denotes the quark-meson coupling constant in scalar and pseudoscalar channels.

To obtain the hadronic version of the topological charge $Q$ and susceptibility $\chi$, we consider $U_A(1)$ transformation for the mesons in scalar and pseudoscalar channels, $\bar\psi\psi\to\bar\psi\psi-2\theta\bar\psi \mathrm{i}\gamma_5\psi/\sqrt 6$ and $\bar\psi \mathrm{i}\gamma_5\psi\to \bar\psi \mathrm{i}\gamma_5\psi+2\theta\bar\psi\psi/\sqrt 6$, which lead to the transformation for the meson matrix $\phi\to (1+2\mathrm{i}\theta/\sqrt 6)\phi$ and $\text{det}(\phi)\to (1+\sqrt 6 \mathrm{i}\theta)\text{det}(\phi)$. By calculating the variation of the Lagrangian density and using the Noether's theorem, the conservation law in the quark-meson model becomes~\cite{Jiang:2015xqz}
\begin{equation}
	\label{j52}
	\partial^\mu J^5_\mu = -12c\text{Im}[\text{det}(\phi)]+2\mathrm{i}\text{Tr}[H(\phi-\phi^\dagger)].
\end{equation}
The second term is due to the explicit chiral symmetry breaking at meson level in the model. Based on the comparison of the first terms in (\ref{j51}) for QCD and (\ref{j52}) for quark-meson model, the topological charge density in the model is as follows:
\begin{equation}
Q(x)=-2c\text{Im}[\text{det}(\phi(x))].
\end{equation}
It contains all possible products of three meson fields.

\begin{figure}[H]
	\centering
	\includegraphics[width=0.4\textwidth]{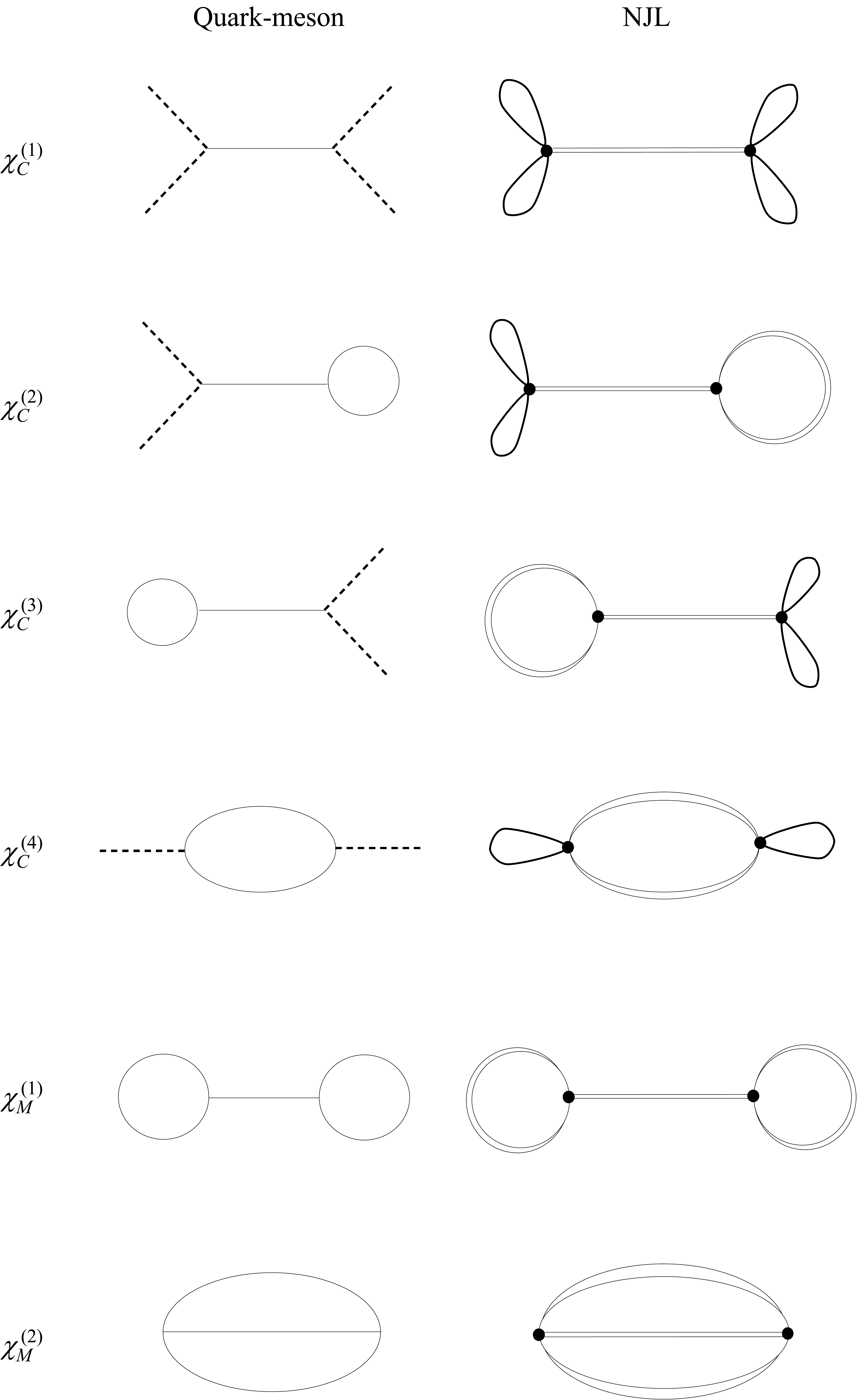}
	\caption{Diagrammatic expression of the topological susceptibility $\chi$ in quark-meson model (left panel) and NJL model (right panel). In the quark-meson model, dashed and solid lines denote chiral condensates and meson propagators, respectively. In the NJL model, the closed propagators at space-time points $0$ or $x$ indicate chiral condensates, and the double lines are meson propagators from $0$ to $x$. }
	\label{fig1}
\end{figure}

We now separate the meson field into a condensate part and a fluctuation part $\phi_a=\langle\phi_a\rangle+\phi_a'$. The former characterizes the spontaneous breaking of the symmetries of the system, and the later is the particle fluctuation above the mean field. Using Wick's theorem, the topological susceptibility (\ref{chi}) consists of the contributions with one, two and three meson propagators between the space-time points $0$ and $x$. The diagram with only condensates is not connected and then neglected. To clearly understand the relation between the chiral symmetry and $U_A(1)$ symmetry, we divide $\chi$ into a sector with chiral condensates and sector with only meson propagators:
\begin{equation}
\chi=\chi_C+\chi_M
\end{equation}
with
\begin{eqnarray}
\chi_C &=& \chi_C^{(1)}+\chi_C^{(2)}+\chi_C^{(3)}+\chi_C^{(4)},\nonumber\\
\chi_C^{(1)} &=& {c^2\over 4} \sum_{abcde}A_{abcde}\langle\phi_a\rangle\langle\phi_b\rangle I_c \langle\phi_d\rangle\langle\phi_e\rangle,\nonumber\\
\chi_C^{(2)} &=& {c^2\over 4} \sum_{abcd}B_{abcd}\langle\phi_a\rangle\langle\phi_b\rangle I_c J_d,\nonumber\\
\chi_C^{(3)} &=& {c^2\over 4} \sum_{abcd}C_{abcd}J_a I_b \langle\phi_c\rangle\langle\phi_d\rangle,\nonumber\\
\chi_C^{(4)} &=& {c^2\over 4} \sum_{abcd}D_{abcd}\langle\phi_a\rangle I_{bc} \langle\phi_d\rangle
\end{eqnarray}
and
\begin{eqnarray}
\chi_M &=& \chi_M^{(1)}+\chi_M^{(2)},\nonumber\\
\chi_M^{(1)} &=& {c^2\over 4} \sum_{abc}E_{abc}J_aI_bJ_c,\nonumber\\
\chi_M^{(2)} &=& {c^2\over 4} \sum_{abc}F_{abc}I_{abc},
\end{eqnarray}
where $A, B, C, D, E$ and $F$ denote the coefficients, $I_a=\int\mathrm{d}^4x G_a(x,0)$, $I_{ab}=\int\mathrm{d}^4x G_a(x,0)G_b(x,0)$ and $I_{abc}=\int\mathrm{d}^4x G_a(x,0)G_b(x,0)G_c(x,0)$ denote the integrated propagator productions with $G_a(x,y)=\langle\phi_a(x)\phi_a(y)\rangle$, and $J_a=G_a(0,0)=G_a(x,x)$ denotes the closed propagator. For simplicity in this expression and subsequent expressions, we replaced the fluctuation field $\phi'$ by $\phi$. $\chi_C^{(i)}\ (i=1,2,3,4)$ and $\chi_M^{(i)}\ (i=1,2)$ which are diagrammatically shown in the left panel of Fig. \ref{fig1}.

Before we analytically and numerically calculate the susceptibility in the next section, we first qualitatively analyze the relation between the chiral and $U_A(1)$ symmetries in chiral limit. In chiral breaking phase at low temperature and density, the chiral condensates and meson degrees of freedom dominate the system wherein the condensate sector $\chi_C\ (\chi_C^{(1)}\sim \langle\phi_a\rangle^4, \chi_C^{(2)}, \chi_C^{(3)}, \chi_C^{(4)}\sim \langle\phi_a\rangle^2)$ and meson sector $\chi_M$ are nonzero, and $U_A(1)$ symmetry is broken. With increasing temperature or baryon chemical potential of the system, the light meson condensate disappears initially at $T_c$ or $\mu_B^c$. However, the strange meson condensate is still nonzero due to the fact that the strange quark is much heavier than the light quarks $m_s\gg m_q$. In this case, the $U_A(1)$ symmetry is still broken as induced by the nonzero $\chi_C$ and $\chi_M$. When the temperature or density increases further with $T\gg T_c$ or $\mu_B\gg \mu_B^c$, the strange meson condensate disappears in the very hot or dense medium, condensate sector $\chi_C$ vanishes completely, and susceptibility is fully controlled by the meson fluctuation part $\chi_M$. In finite-temperature field theory, a Feynman diagram with a particle loop contributes a factor of particle number distribution $n$ (Bose-Einstein distribution $n_B$ or Fermi-Dirac distribution $n_F$). Please refer to any textbook, such as Ref.~\cite{Kapusta:2006pm}, for more details. Furthermore, detailed calculations are provided in the next section. For the Feynman diagrams in $\chi_M$ shown in Fig. \ref{fig1}, a meson loop, corresponding to a gluon loop in QCD, contributes a Bose-Einstein distribution $n_B(\epsilon_a^p)=1/(\mathrm{e}^{\epsilon_a^p/T}-1)$ with meson energy $\epsilon_a^p=\sqrt{m_a^2+{\bm p}^2}$. It should be noted that the quark chemical potential $(\mu_B/3)$ does not enter the quark-antiquark pair distribution. At zero temperature, there is no thermal excitation of mesons ($n_B=0$), and therefore the meson sector $\chi_M$ disappears. This implies that, the $U_A(1)$ symmetry can be restored strictly only by the density effect at zero temperature.

The aforementioned conclusion also applies to the NJL model at quark level. In the three-flavor NJL model~\cite{Klevansky:1992qe}, the $U_A(1)$ symmetry is broken by a six-quark interaction with a coupling constant $K$. Under the $U_A(1)$ transformation, the topological charge can be directly derived~\cite{Fukushima:2001hr}
\begin{equation}
Q(x) = 2K\text{Im}\ \text{det}\left[\bar\psi(x)(1-\gamma_5)\psi(x)\right]
\end{equation}
with all possible products of six quark fields at space-time point $x$. The corresponding Feynman diagrams for the condensate sector $\chi_C^{(i)}$ and meson sector $\chi_M^{(i)}$ of the susceptibility $\chi$ are shown in the right panel of Fig. \ref{fig1}. In comparison with the left panel, the diagrams in the NJL model are very similar to that in the quark-meson model: the meson condensates $\langle\phi_a\rangle$ (dashed lines) now become the quark-antiquark condensate $\langle\bar q q\rangle$ and $\langle\bar s s\rangle$ (closed quark propagators at $0$ or $x$), and the mesons (solid lines) are constructed by quarks via random phase approximation~\cite{Klevansky:1992qe} at order $\mathcal O(1/N_c)$ (double lines). Given that the susceptibilities in the two models have the same structure, we again conclude that the $U_A(1)$ symmetry breaking can only be restored by pure baryon density effect. The detailed calculation on the $U_A(1)$ symmetry at finite temperature in the NJL-type model can be seen in Ref.\cite{Fukushima:2001hr,Fukushima:2001ut,Costa:2002gk,Costa:2003uu,Costa:2004db,Costa:2005cz,Chen:2009gv,Brauner:2009gu,Contrera:2009hk,Ruivo:2011fg,Ruivo:2012xt,Xia:2013caa,Cui:2021bqf}.

\section{Analytic and numerical calculations}
\label{s3}
In this section, we analytically and numerically calculate the topological susceptibility in the quark-meson model at finite baryon density. We will address calculations in the real world that involve explicit chiral symmetry breaking. Given that the susceptibility is dependent on the condensates, as well as meson and quark masses, we will first provide a brief overview of the condensates and masses using the mean field approximation. Detailed calculations can be sourced from existing literature \cite{Schaefer:2008hk}.

\subsection{Condensates and masses}
After the separation of the meson field into a condensate part and a fluctuation part $\phi=\langle\phi\rangle+\phi'$, a meson potential $V_M(\langle\phi\rangle)$~\cite{Lenaghan:2000ey,Schaefer:2008hk} appears in the Lagrangian $\mathcal L_M$. At mean field level, it is the thermodynamic potential of the system $\Omega_M=V_M$. Considering the thermodynamics from the free constituent quarks with mass:
\begin{equation}
m=gT^a\left(\langle\sigma_a\rangle+\mathrm{i}\gamma_5\langle\pi_a\rangle\right),
\end{equation}
the thermodynamic potential of the quark-meson system becomes
\begin{equation}
\Omega=\Omega_M+\Omega_Q
\end{equation}
with
\begin{equation}
\Omega_Q = 2N_cT\sum_f\int \frac{\mathrm{d}^3 {\bm p}}{(2\pi)^3}\left[\ln\left(1-n_F(\epsilon_f^p)\right)+\ln\left(1-\bar n_F(\epsilon_f^p)\right)\right],
\end{equation}
where $n_F=1/(\mathrm{e}^{(\epsilon_f^p-\mu_B/3)/T}+1)$ and $\bar n_F=1/(\mathrm{e}^{(\epsilon_f^p+\mu_B/3)/T}+1)$ denote the Fermi-Dirac distributions for constituent quarks and anti-quarks, and $\epsilon_f^p=\sqrt{m_f^2+{\bm p}^2}$ denotes the quark energy with flavor $f$.

The physical condensates as functions of temperature and baryon chemical potential $\langle\phi_a\rangle(T,\mu_B)$ are determined by minimizing the thermodynamic potential
\begin{equation}
\label{gap}
{\partial\Omega\over \partial \langle\phi_a\rangle} =0,\quad\quad {\partial^2\Omega\over \partial \langle\phi_a\rangle^2}>0.
\end{equation}

In mean field approximation, the meson masses can be directly derived from the quadratic term in the Lagrangian $\tilde m_a^2=\partial^2\mathcal L/\partial\phi_a^2|_{\phi=0}$, which is equivalent to the second coefficient of the Taylor expansion of $\Omega_M(\langle\phi\rangle)$ around the physical condensate determined by the gap equation (\ref{gap}), $\tilde m_a^2=\partial^2\Omega_M/\partial\langle\phi_a\rangle^2$. To contain the contribution from quark thermodynamics to meson masses, one phenomenological approach to go beyond the mean field is to extend the second order derivative from $\Omega_M$ to the total potential $\Omega$~\cite{Schaefer:2008hk},
\begin{equation}
m_a^2={\partial^2\Omega\over \partial\langle\phi_a\rangle^2}=\tilde m_a^2+{\partial^2\Omega_Q\over\partial\langle\phi_a\rangle^2}.
\end{equation}

Given that we focus on the chiral symmetry and $U_A(1)$ symmetry in this study, we introduce only the chiral condensates $\langle\sigma_0\rangle$ and $\langle\sigma_8\rangle$ in the following. Considering the mixing between $\phi_0$ and $\phi_8$, normally a rotation in this subspace is considered. The two condensates are changed to the chiral condensate $\langle\sigma\rangle_c= 1/\sqrt 3(\sqrt 2\langle\sigma_0\rangle+\langle\sigma_8\rangle)$ and strange condensate $\langle\sigma\rangle_s=1/\sqrt 3(\langle\sigma_0\rangle-\sqrt 2\langle\sigma_8\rangle)$, which leads to the constituent mass $m_q=g\langle\sigma\rangle_c/2$ for light quarks and $m_s=g\langle\sigma\rangle_s/\sqrt 2$ for strange quarks. In the pseudoscalar channel, $\pi_0$ and $\pi_8$ are rotated to the experimentally measured mesons $\eta$ and $\eta'$ via $\pi_0=\cos\theta_p\eta'-\sin\theta_p\eta$ and $\pi_8=\sin\theta_p\eta'+\cos\theta_p\eta$ with the rotation angle $\theta_p$.

With the choice of condensates and under the rotation, the four independent pseudoscalar meson masses in mean field approximation, $\tilde m_\pi^2$ for $a=1,2,3$, $\tilde m_K^2$ for $a=4,5,6,7$, $\tilde m_{\eta'}^2$ and $\tilde m_\eta^2$, can be explicitly expressed in terms of the chiral and strange condensates,
\begin{eqnarray}
\tilde m_\pi^2 &=& \lambda^2+\lambda_1\left(\langle\sigma\rangle_c^2+\langle\sigma\rangle_s^2\right)+\frac{\lambda_2}{2} \langle\sigma\rangle_c^2-\frac{c}{\sqrt 2} \langle\sigma\rangle_s,\\
\tilde m_K^2 &=& \lambda^2+\lambda_1\left(\langle\sigma\rangle_c^2+\langle\sigma\rangle_s^2\right)\nonumber\\
&&+\frac{\lambda_2}{2}\left(\langle\sigma\rangle_c^2-\sqrt 2\langle\sigma\rangle_c \langle\sigma\rangle_s +2\langle\sigma\rangle_s^2\right) -\frac{c}{2}\langle\sigma\rangle_c,\nonumber\\
\tilde m_{\eta'}^2 &=& m_{00}^2\cos^2\theta_p+m_{88}^2\sin^2\theta_p+2m_{08}^2\sin\theta_p\cos\theta_p,\nonumber\\
\tilde m_\eta^2 &=& m_{00}^2\sin^2\theta_p+m_{88}^2\cos^2\theta_p-2m_{08}^2\sin\theta_p\cos\theta_p\nonumber
\end{eqnarray}
with
\begin{eqnarray}
m_{00}^2 &=& \lambda^2+\lambda_1\left(\langle\sigma\rangle_c^2+\langle\sigma\rangle_s^2\right)+{\lambda_2\over 3}\left(\langle\sigma\rangle_c^2+\langle\sigma\rangle_s^2\right)\\
&& +\frac{\sqrt 2c}{3}\left(\sqrt 2\langle\sigma\rangle_c+\langle\sigma\rangle_s\right),\nonumber\\
m_{88}^2 &=& \lambda^2+\lambda_1\left(\langle\sigma\rangle_c^2+\langle\sigma\rangle_s^2\right)+{\lambda_2\over 6}\left(\langle\sigma\rangle_c^2+4\langle\sigma\rangle_s^2\right)\nonumber\\
&& -\frac{\sqrt 2c}{6}\left(2\sqrt 2\langle\sigma\rangle_c-\langle\sigma\rangle_s\right),\nonumber\\
m_{08}^2 &=& \frac{\sqrt 2\lambda_2}{6}\left(\langle\sigma\rangle_c^2-2\langle\sigma\rangle_s^2\right)-\frac{\sqrt 2c}{6}\left(\langle\sigma\rangle_c-\sqrt 2\langle\sigma\rangle_s\right)\nonumber
\end{eqnarray}
and the mixing angle $\tan 2\theta_p = 2m_{08}^2/(m_{00}^2-m_{88}^2)$. Similarly, we can obtain the scalar meson masses~\cite{Schaefer:2008hk} $m_{a_0}^2, m_\kappa^2, m_\sigma^2$ and $m_{f_0}^2$.

The model parameters $\lambda^2, \lambda_1, \lambda_2, h_c, h_s, c$ and $g$ and the condensates $\langle\sigma\rangle_c$ and $\langle\sigma\rangle_s$ in vacuum should to be fixed by fitting the meson properties in vacuum. By choosing the pseudoscalar meson masses $m_\pi=135$ MeV, $m_K=496$ MeV, $m_\eta=539$ MeV and $m_{\eta'}=963$ MeV and the decay constants $f_\pi=92.4$ MeV and $f_K=113$ MeV~\cite{ParticleDataGroup:2020ssz}, we can determine six of them, namely the meson coupling constant $\lambda_2$,
\begin{eqnarray}
\lambda_2 &=& \frac{3\left(2f_K-f_\pi\right)m_K^2-\left(2f_K^2+f_\pi\right)m_\pi^2-2\left(m_\eta^2+m_{\eta^\prime}^2\right)\left(f_K-f_\pi\right)}
{\left[3f_\pi^2+8f_K\left(f_K-f_\pi\right)\right]\left(f_K-f_\pi\right)}\nonumber\\
&=& 46.4881,
\end{eqnarray}
parameter $c$ controlling $U_A\left(1\right)$ symmetry breaking,
\begin{equation}
c=\frac{m_K^2-m_\pi^2}{f_K-f_\pi}-\lambda_2\left(2f_K-f_\pi\right)=4807.24~\text{MeV},
\end{equation}
parameters $h_c$ and $h_s$ governing chiral symmetry breaking,
\begin{eqnarray}
h_c &=& f_\pi m_\pi^2 = (120.729\ \text{MeV})^3,\nonumber\\
h_s &=& \sqrt{2}f_Km_K^2-\frac{f_\pi m_\pi^2}{\sqrt{2}} = (336.406\ \text{MeV})^3,
\end{eqnarray}
and chiral condensates $\langle\sigma\rangle_c$ and $\langle\sigma\rangle_s$,
\begin{eqnarray}
\langle\sigma\rangle_c &=& f_\pi = 92.4\ \text{MeV},\nonumber\\
\langle\sigma\rangle_s &=& \frac{1}{\sqrt{2}}\left(2f_K-f_\pi\right) = 94.48\ \text{MeV}.
\end{eqnarray}
To determine the other meson coupling $\lambda_1$ and mass parameter $\lambda^2$, scalar mesons are required. Considering $m_\sigma=550$ MeV, we obtain $\lambda^2=(393.945\ \text{MeV})^2$ and $\lambda_1=-0.771779$. The quark-meson coupling $g$ and strange quark mass $m_s$ are further associated with the non-strange quark mass $m_q$. By choosing $m_q=300$ MeV, we obtain $g = 6.4$ and $m_s=433$ MeV.
\begin{figure}[H]
	\centering
	\includegraphics[width=0.35\textwidth]{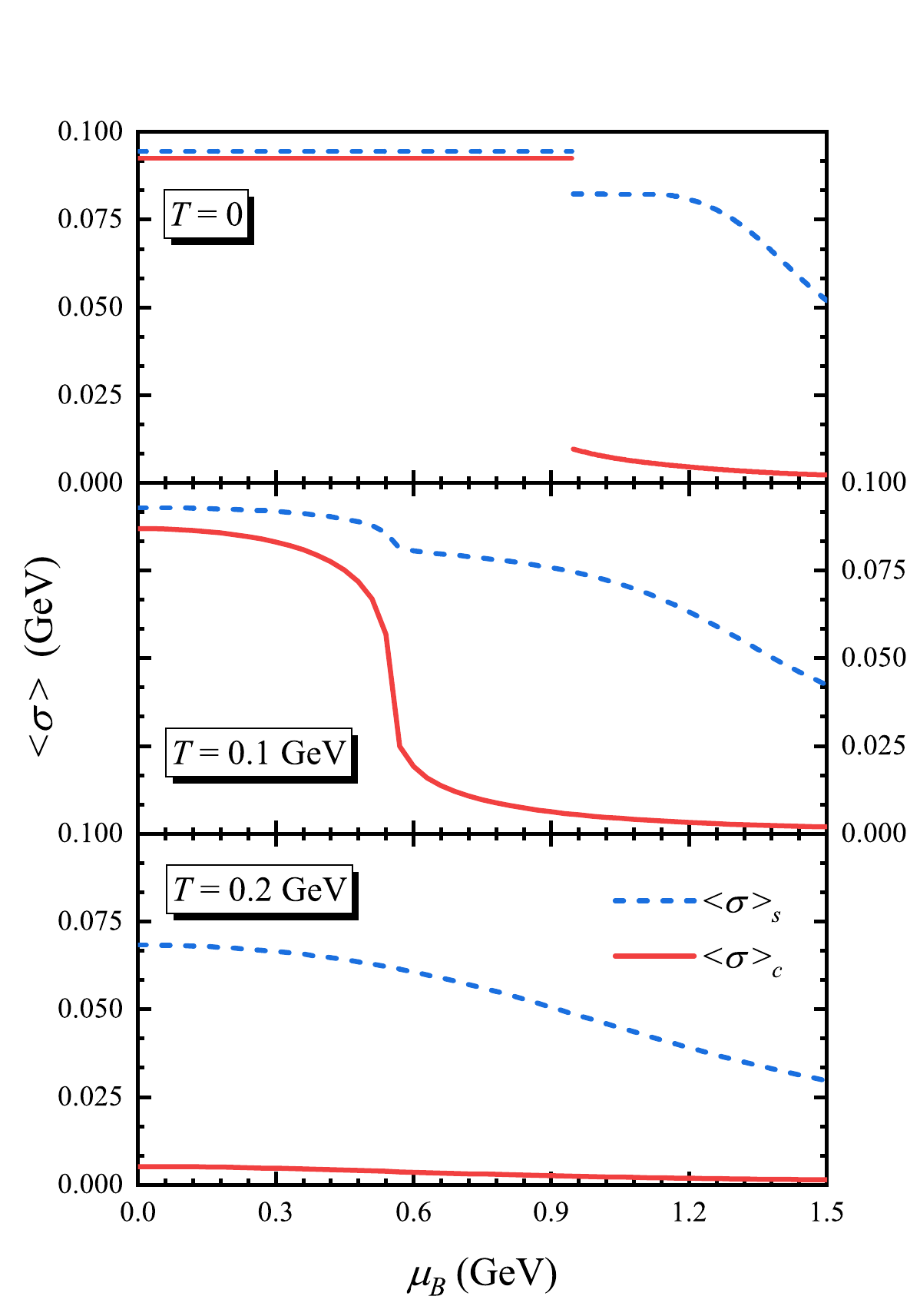}
	\caption{Chiral and strange condensates $\langle\sigma\rangle_c$ (solid lines) and $\langle\sigma\rangle_s$ (dashed lines) as functions of baryon chemical potential $\mu_B$ at temperature $T$=0 (upper panel), 0.1 (middle panel) and 0.2 (lower panel) GeV. }
	\label{fig2}
\end{figure}

As the $1/N_c$ realization of the t'Hooft instanton mechanism, the Witten--Veneziano (WV) formula~\cite{Witten:1979vv,Veneziano:1979ec} is as follows:
\begin{equation}
\label{wv}
\chi_{\text{pure}} = {m_\eta^2+m_{\eta'}^2-2m_K^2\over 2N_f}f_\pi^2 + \mathcal{O}\left(\frac{1}{N_c}\right).
\end{equation}
This can be applied to estimate the $U_A(1)$ symmetry breaking in vacuum through the pseudoscalar meson masses and pion decay constant (It should be noted that the susceptibility in the WV formula is for the pure Yang-Mills theory). The formula is confirmed by effective methods~\cite{Fukushima:2001hr,Fukushima:2001ut} and lattice QCD calculations~\cite{DelDebbio:2004ns,Durr:2006ky,Cichy:2015jra}. In our calculation, the above used parameters leads to $\chi=(191.033 \text{MeV})^4$ which is in good agreement with the lattice result $\chi=(191\pm 5 \text{MeV})^4$ in continuum limit~\cite{DelDebbio:2004ns}. However, it is claimed that the WV formula cannot be extended to finite temperature, especially near the QCD critical point~\cite{Horvatic:2007qs,Benic:2011fv}.

With the known parameters, we now numerically calculate the density and temperature dependence of the two scalar condensates, and the result is shown in Fig. \ref{fig2}. Governed by the Fermi surface at zero temperature, the chiral condensate retains its vacuum value at low densities and then abruptly drops to a significantly lower value upon reaching the critical chemical potential $\mu_B^c=0.91$ GeV, and then decreases smoothly. For the strange condensate, there is also a jump at $\mu_B^c$, but it is still large in the chiral restoration phase. As the temperature increases, the abrupt changes in the two condensates gradually diminish, transitioning the chiral phase from a distinct jump to a crossover. In sufficiently hot conditions, this crossover occurs at zero baryon density.

\begin{figure}[H]
	\centering
	\includegraphics[width=0.35\textwidth]{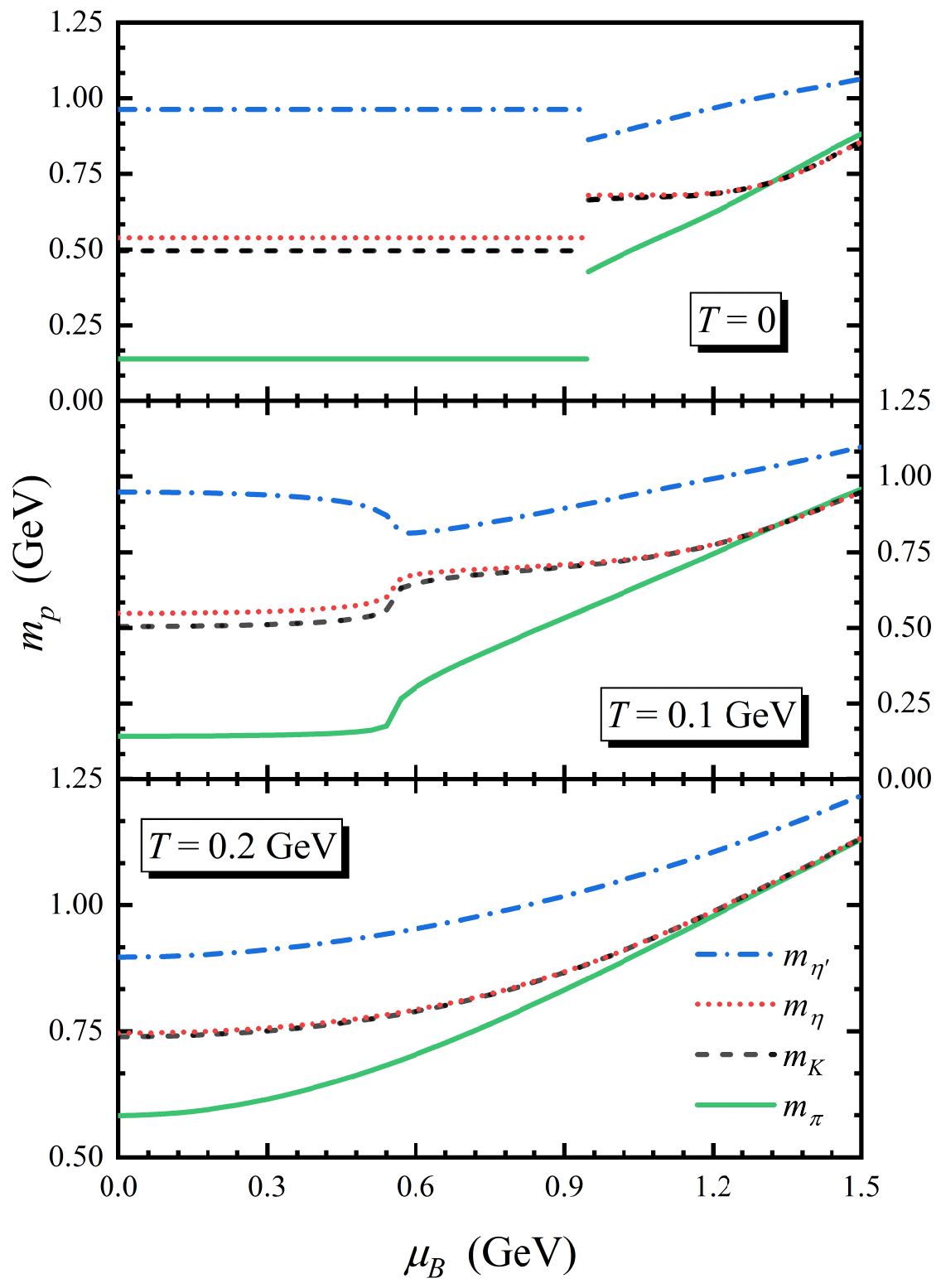}
	\caption{Pseudoscalar meson masses $m_\pi$ (solid lines), $m_K$ (dashed lines), $m_\eta$(dotted lines) and $m_{\eta'}$ (dot-dashed lines) as functions of baryon chemical potential $\mu_B$ at temperature $T$=0 (upper panel), 0.1 (middle panel) and 0.2 (lower panel) GeV. }
	\label{fig3}
\end{figure}

The density and temperature dependence of the pseudoscalar meson masses is shown in Fig. \ref{fig3}. At zero temperature, all the masses consistently retain their vacuum values below the critical chemical potential. However, they abruptly increase or decrease at the chiral phase transition point $\mu_B^c$ and change continuously afterwards. The strange meson $K$ is heavier than the pseudo-Goldstone particle $\pi$ in the chiral breaking phase at low density. However, the two masses approach each other in the chiral restoration phase when the chemical potential is larger than the strange quark mass. The large mass splitting between $\eta$ and $\eta'$ at low density is induced by the $U_A(1)$ breaking. At the critical point, $m_\eta$ experiences an upward shift while $m_{\eta'}$ decreases, and the disparity between them decreases as density increases. When the temperature effect is included, all the jumps will gradually be replaced by continuous changes.

\subsection {Susceptibility}
The susceptibility $\chi$ varies based on density and temperature, influenced by the condensates, meson masses, and the loop induced Bose-Einstein distribution $n_B$. We first calculate the four independent meson constituents shown in Fig. \ref{fig1}, namely the closed meson propagator $J_a$, tmeson propagator with zero momentum $I_a$, meson loop constructed by two mesons $I_{ab}$ and double meson loops by three mesons $I_{abc}$,
\begin{eqnarray}
&& J_a = \int {d^3{\bm p}\over (2\pi)^3}{n_B(\epsilon_a^p)\over \epsilon_a^p},\nonumber\\
&& I_a = \frac{1}{m_a^2},\nonumber\\
&& I_{ab} = \int{d^3{\bm p}\over (2\pi)^3}\frac{1}{m_b^2-m_a^2}\left[\frac{n_B(\epsilon_a^p)}{\epsilon_a^p}-\frac{n_B(\epsilon_b^p)}{\epsilon_b^p}\right],\nonumber\\
&& I_{abc} = I_{abc}^{(1)}+I_{abc}^{(2)}.
\end{eqnarray}
The term $I_{abc}$ contains two four-momentum integrations (two Matsubara frequency summations and two three-momentum integrations), and each frequency summation contributes a constant and meson distribution $n_B$. After considering a renormalization to remove the divergence appeared in the $n_B$-independent integration~\cite{Baacke:2002pi}, $I_{abc}$ is separated into a part $I_{abc}^{(1)}$ with one meson distribution and a part $I_{abc}^{(2)}$ with two distributions,
\begin{eqnarray}
I_{abc}^{(1)} &=& {1\over (4\pi)^2}\sum_{\{abc\}}\Bigg\{-\gamma_E+\ln(4\pi)-\ln\left({m_c^2\over \mu^2}\right)-\int_0^1 d\alpha\Bigg[\alpha{m_a^2\over m_c^2}\nonumber\\
&& +(1-\alpha){m_b^2\over m_c^2}-\alpha(1-\alpha)\Bigg]\Bigg\}\int{d^3{\bm p}\over (2\pi)^3}{n_B(\epsilon_c^p)\over \epsilon_c^p},\nonumber\\
I_{abc}^{(2)} &=& 2\pi^2\sum_{\{abc\}}\int{d^3{\bm p}\over (2\pi)^3}{d^3{\bm q}\over (2\pi)^3}{{\bm p}\cdot{\bm q}n_B(\epsilon_a^p)n_B(\epsilon_b^q)\over \epsilon_a^p\epsilon_b^q}\nonumber\\
&&\times\ln\left|{\left(\epsilon_a^p+\epsilon_b^q\right)^2-\left(\epsilon_c^{p+q}\right)^2\over \left(\epsilon_a^p+\epsilon_b^q\right)^2-\left(\epsilon_c^{p-q}\right)^2}{\left(\epsilon_a^p-\epsilon_b^q\right)^2-\left(\epsilon_c^{p+q}\right)^2\over \left(\epsilon_a^p-\epsilon_b^q\right)^2-\left(\epsilon_c^{p-q}\right)^2}\right|,
\end{eqnarray}
where $\gamma_E$ denote the Euler constant, the renormalization scale $\mu$ is considered to be 0.3 GeV in the calculation, and the sum is defined as $\sum_{\{abc\}}X_{abc}=X_{abc}+X_{bca}+X_{cab}$. 

With the mixing angels $\theta_s$ and $\theta_p$ in the scalar and pseudiscalar channels, we define the diagonalization coefficients as follows:
\begin{eqnarray}
c_1 &=& \sqrt{1/3}\left(\cos\theta_s-\sqrt 2\sin\theta_s\right),\nonumber\\
c_2 &=& \sqrt{1/3}\left(\sin\theta_s+\sqrt 2\cos\theta_s\right),\nonumber\\
c_3 &=& \sqrt{1/3}\left(\cos\theta_p-\sqrt 2\sin\theta_p\right),\nonumber\\
c_4 &=& \sqrt{1/3}\left(\sin\theta_p+\sqrt 2\cos\theta_p\right),
\end{eqnarray}
the vertexes of the Feynman diagrams in Fig. \ref{fig1} can then be expressed as
\begin{eqnarray}
c_{\eta\sigma\langle\sigma\rangle_c} &=&\sqrt 2/3 \left(2\sqrt 2\sin(\theta_p+\theta_s)+\cos(\theta_p+\theta_s)\right),\nonumber\\
c_{\eta f_0\langle\sigma\rangle_c} &=& \sqrt 2/3 \left(2\sqrt{2}\cos(\theta_p+\theta_s)-\sin(\theta_p+\theta_s)\right),\nonumber\\
c_{\eta'\sigma\langle\sigma\rangle_c} &=& -c_{\eta f_0\langle\sigma\rangle_c},\nonumber\\
c_{\eta'f_0\langle\sigma\rangle_c} &=& c_{\eta\sigma\langle\sigma\rangle_c},\nonumber\\
c_{\eta\sigma\langle\sigma\rangle_s} &=& \sqrt 2/6\Big(3\sin(\theta_p-\theta_s)-2\sqrt 2\cos(\theta_p+\theta_s)\nonumber\\
&&+\sin(\theta_p+\theta_s)\Big),\nonumber\\
c_{\eta f_0\langle\sigma\rangle_s} &=&-\sqrt 2/6\Big(3\cos(\theta_p-\theta_s)-2\sqrt 2\sin(\theta_p+\theta_s)\nonumber\\
&& -\cos(\theta_p+\theta_s)\Big),\nonumber\\
c_{\eta'\sigma\langle\sigma\rangle_s} &=& -\left(c_{\eta f_0\langle\sigma\rangle_s}+\sqrt 2\cos(\theta_p-\theta_s)\right),\nonumber\\
c_{\eta' f_0\langle\sigma\rangle_s} &=& c_{\eta\sigma\langle\sigma\rangle_s}-\sqrt 2\sin(\theta_p-\theta_s)
\end{eqnarray}
for the vertexes with one condensate leg, and
\begin{eqnarray}
c_{\eta\eta\eta} &=& -1/\sqrt 2c_3^2c_4,\nonumber\\
c_{\eta\eta\eta'} &=& 1/\sqrt 2\left(c_3^3-2c_3c_4^2\right),\nonumber\\
c_{\eta\eta'\eta'} &=& 1/\sqrt 2\left(2c_3^2c_4-c_4^3\right),\nonumber\\
c_{\eta'\eta'\eta'} &=& 1/\sqrt 2c_3c_4^2,\nonumber\\
c_{\eta\sigma\sigma} &=& 1/\sqrt 3\Big(\cos\theta_p(\sin^2\theta_s+1/\sqrt 2\sin 2\theta_s)\nonumber\\
&&+\sqrt 2\sin\theta_p(\cos^2\theta_s-1/2\sin^2\theta_s)\Big),\nonumber\\
c_{\eta\sigma f_0} &=& 1/\sqrt 3\Big(\cos\theta_p(\sin 2\theta_s+\sqrt 2\cos2\theta_s)\nonumber\\
&&-3/\sqrt 2\sin\theta_p\sin2\theta_s\Big),\nonumber\\
c_{\eta f_0f_0} &=& 1/\sqrt 3\Big(\cos\theta_p(\cos^2\theta_s-1/\sqrt 2\sin 2\theta_s)\nonumber\\
&&+\sqrt 2\sin\theta_p(\sin^2\theta_s-1/2\cos^2\theta_s)\Big),\nonumber\\
c_{\eta'\sigma\sigma} &=& 1/\sqrt 3\Big(\sin\theta_p(\sin^2\theta_s+1/\sqrt 2\sin 2\theta_s)\nonumber\\
&& -\sqrt 2\cos\theta_p(\cos^2\theta_s-1/2\sin^2\theta_s)\Big),\nonumber\\
c_{\eta'\sigma f_0} &=& 1/\sqrt 3\Big(\sin\theta_p(\sin2\theta_s+\sqrt 2\cos2\theta_s)\nonumber\\
&& +3/\sqrt 2\cos\theta_p\sin2\theta_s\Big),\nonumber\\
c_{\eta'f_0f_0} &=& 1/\sqrt 3\Big(\sin\theta_p(\cos^2\theta_s-1/\sqrt 2\sin 2\theta_2)\nonumber\\
&& -\sqrt 2\cos\theta_p(\sin^2\theta_s-1/2\cos^2\theta_s)\Big)
\end{eqnarray}
for the vertexes without condensate legs. Finally, we define two new condensates
\begin{eqnarray}
\langle\sigma^2_\eta\rangle &=& 1/\sqrt 6\langle\sigma\rangle_c\Big((\sqrt 2\cos\theta_p+\sin\theta_p)\langle\sigma\rangle_c\nonumber\\
&& +2(\sqrt 2\sin\theta_p-\cos\theta_p)\langle\sigma\rangle_s\Big),\nonumber\\
\langle\sigma^2_{\eta'}\rangle &=& 1/\sqrt{6}\langle\sigma\rangle_c\Big((\sqrt 2\sin\theta_p-\cos\theta_p)\langle\sigma\rangle_c\nonumber\\
&& -2(\sqrt 2\cos\theta_p+\sin\theta_p)\langle\sigma\rangle_s\Big),
\end{eqnarray}
and explicitly write the different susceptibility terms:
\begin{widetext}
\begin{eqnarray}
\chi_C^{(1)} &=& {c^2\over 4}\Big[\langle\sigma^2_\eta\rangle^2I_\eta+\langle\sigma^2_{\eta'}\rangle^2I_{\eta'}\Big],\nonumber\\
\chi_C^{(2)} &=& {c^2\over 4}\Big[\langle\sigma^2_\eta\rangle\Big(6c_{\eta\eta\eta}J_\eta+2c_{\eta\eta'\eta'}J_{\eta'}+2c_{\eta\sigma\sigma}J_\sigma+2c_{\eta f_0 f_0}J_{f_0}+4c_3(J_\kappa-J_K)+3\sqrt 2c_4(J_\pi-J_{a_0})\Big)I_\eta\nonumber\\
&&+\langle\sigma^2_{\eta'}\rangle\Big(6c_{\eta'\eta'\eta'}J_{\eta'}+2c_{\eta\eta\eta'}J_\eta+2c_{\eta'\sigma\sigma}J_\sigma+2c_{\eta' f_0 f_0}J_{f_0}+3\sqrt 2 c_3(J_{a_0}-J_\pi)+4c_4(J_\kappa-J_K)
\Big)I_{\eta'}\Big],\nonumber\\
\chi_C^{(3)} &=& \chi_C^{(2)},\nonumber\\
\chi_C^{(4)} &=& {c^2\over 4}\Big[(c_{\eta\sigma\langle\sigma\rangle_c}\langle\sigma\rangle_c+c_{\eta\sigma\langle\sigma\rangle_s}\langle\sigma\rangle_s)^2 I_{\eta\sigma}+(c_{\eta f_0\langle\sigma\rangle_c}\langle\sigma\rangle_c+c_{\eta f_0\langle\sigma\rangle_s}\langle\sigma\rangle_s)^2I_{\eta f_0}+(c_{\eta'\sigma\langle\sigma\rangle_c}\langle\sigma\rangle_c+c_{\eta'\sigma\langle\sigma\rangle_s}\langle\sigma\rangle_s)^2I_{\eta'\sigma}\nonumber\\
&&+(c_{\eta' f_0\langle\sigma\rangle_c}\langle\sigma\rangle_c+c_{\eta' f_0\langle\sigma\rangle_s}\langle\sigma\rangle_s)^2 I_{\eta'f_0}
+4\langle\sigma\rangle_c^2I_{K\kappa}+6\langle\sigma\rangle_s^2 I_{\pi a_0}\Big]
\end{eqnarray}
for the condensate controlled part $\chi_C$, and
\begin{eqnarray}
\chi_M^{(1)} &=& {c^2\over 4}\Big\{I_\eta\Big[3c_{\eta\eta\eta}\Big(3c_{\eta\eta\eta} J_\eta+2c_{\eta\eta'\eta'} J_{\eta'}+3 \sqrt 2 c_4 (J_\pi-J_{a_0})+4 c_3(J_\kappa-J_K)+2(c_{\eta\sigma\sigma}J_\sigma+c_{\eta f_0 f_0}J_{f_0})\Big)J_\eta\nonumber\\
&&+c_{\eta\eta'\eta'}\Big(c_{\eta\eta'\eta'}J_{\eta'}+2 c_{\eta\sigma\sigma}J_\sigma+2c_{\eta f_0 f_0}J_{f_0}+3\sqrt 2c_4(J_\pi-J_{a_0})+4c_3(J_\kappa-J_K)\Big)J_{\eta'}\nonumber\\
&&+c_{\eta\sigma\sigma}\Big(c_{\eta\sigma\sigma}J_\sigma+2c_{\eta f_0 f_0}J_{f_0}+3\sqrt{2}c_4(J_\pi-J_{a_0})+4c_3(J_\kappa-J_K)\Big)J_\sigma+c_{\eta f_0 f_0}\Big(c_{\eta f_0 f_0} J_{f_0}+3\sqrt{2}c_4(J_\pi-J_{a_0})+4c_3(J_\kappa-J_K)\Big)J_{f_0}\nonumber\\
&&+3/2c_4^2\Big(3J_{a_0}J_{a_0}+3J_\pi J_\pi+2J_\kappa J_\kappa-6J_\pi J_{a_0}\Big)+c_3^2\Big(4J_K J_K+J_\kappa J_\kappa-8J_K J_\kappa\Big)+6\sqrt 2c_3c_4\Big(J_{a_0}J_K-J_{a_0}J_\kappa-J_\pi J_K+J_\pi J_\kappa\Big)\Big]\nonumber\\
&&+I_{\eta'}\Big[3c_{\eta'\eta'\eta'}\Big(3c_{\eta'\eta'\eta'}J_{\eta'}+2 c_{\eta'\eta\eta} J_\eta +3\sqrt 2 c_3(J_{a_0}-J_\pi)+4 c_4 (J_\kappa-J_K)+2(c_{\eta'\sigma\sigma}J_\sigma+c_{\eta' f_0 f_0}J_{f_0})\Big)J_{\eta'}\nonumber\\
&&+c_{\eta\eta\eta'}\Big(2c_{\eta'\sigma\sigma} J_\sigma+2 c_{\eta'f_0 f_0} J_{f_0}+3 \sqrt 2 c_3(J_{a_0}-J_\pi)+4c_4(J_\kappa-J_K) +c_{\eta\eta\eta'} J_\eta\Big)J_\eta\nonumber\\
&&+c_{\eta'\sigma\sigma}\Big(c_{\eta'\sigma\sigma}J_\sigma+2c_{\eta' f_0 f_0}J_{f_0}+3\sqrt 2c_3(J_{a_0}-J_\pi)+4c_4(J_\kappa-J_K)\Big)J_\sigma+c_{\eta' f_0 f_0}\Big(c_{\eta' f_0 f_0}J_{f_0}+3\sqrt 2 c_3(J_{a_0}-J_\pi)+4c_4(J_\kappa-J_K)\Big)J_{f_0}\nonumber\\
&&+c_4^2\Big(4J_K J_K+3J_\kappa J_\kappa-8J_K J_\kappa\Big)+9/2c_3^2\Big(J_{a_0}J_{a_0}+J_\pi J_\pi-2J_\pi J_{a_0}\Big)+6\sqrt 2c_3c_4\Big(J_\pi J_K-J_\pi J_\kappa-J_{a_0}J_K-J_{a_0}J_\kappa\Big)\Big]\Big\},\nonumber\\
\chi_M^{(2)} &=& {c^2\over 4}\Big[6I_{\pi K K}+6I_{\pi \kappa \kappa}+12I_{K a_0 \kappa}+6c^2_{\eta\eta\eta}I_{\eta \eta \eta}+2c^2_{\eta\eta\eta'}I_{\eta \eta \eta'}+6c^2_{\eta'\eta'\eta'}I_{\eta' \eta' \eta'}+2c^2_{\eta\eta'\eta'}I_{\eta\eta'\eta'}+2c^2_{\eta'\sigma\sigma}I_{\eta' \sigma\sigma}+2c^2_{\eta' f_0 f_0}I_{\eta'f_0f_0}\nonumber\\
&&+c^2_{\eta'\sigma f_0}I_{\eta'\sigma f_0}+c^2_{\eta\sigma f_0}I_{\eta\sigma f_0}+2c^2_{\eta\sigma\sigma}I_{\eta\sigma\sigma}+2c^2_{\eta f_0 f_0}I_{\eta f_0 f_0}+6c_1^2I_{\sigma\pi a_0}+6c_2^2I_{f_0\pi a_0}+4c_1^2I_{f_0 K \kappa}+4c_2^2I_{\sigma K\kappa}+3c_3^2I_{\eta' a_0 a_0}\nonumber\\
&&+3c_3^2I_{\eta'\pi\pi}+2c_4^2I_{\eta' K K}+3c_4^2I_{\eta'\kappa\kappa}+3c_4^2I_{\eta a_0 a_0}+3c_4^2I_{\eta\pi\pi}+2c_3^2I_{\eta K K}+3c_3^2I_{\eta\kappa\kappa}\Big]
\end{eqnarray}
\end{widetext}
for the meson fluctuation controlled part $\chi_M$.

\begin{figure}[H]
	\centering
	\includegraphics[width=0.35\textwidth]{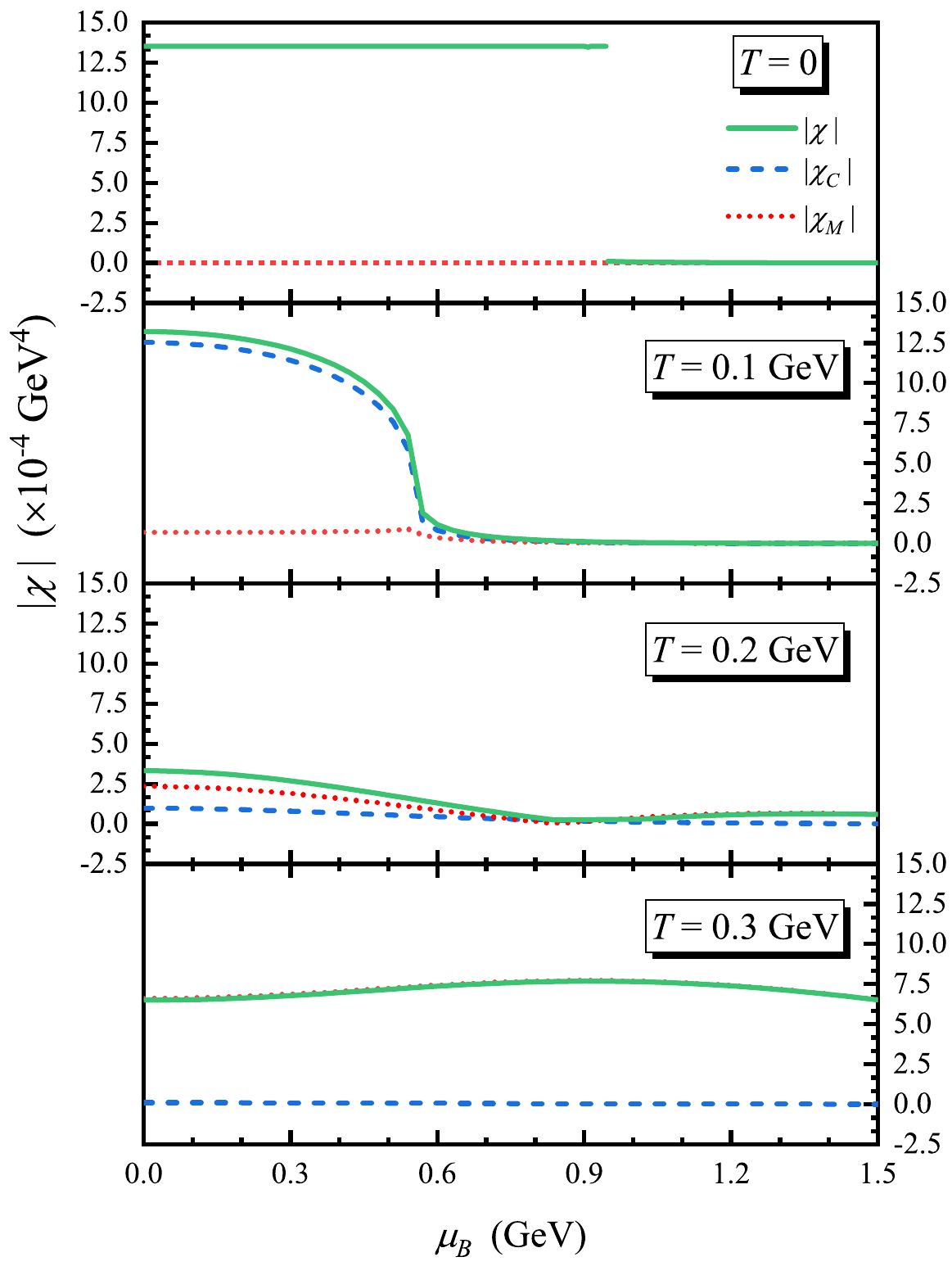}
	\caption{Absolute values of topological susceptibility $\chi$ (solid lines) and its condensate component $\chi_C$ (dashed lines) and meson component $\chi_M$ (dotted lines) as functions of baryon chemical potential $\mu_B$ at temperature $T=0,\ 0.1,\ 0.2$ and $0.3$ GeV. }
	\label{fig4}
\end{figure}

The topological susceptibility $\chi$ and its two components $\chi_C$ and $\chi_M$ in dense and hot quark-meson matter are shown in Fig. \ref{fig4}. To clearly observe the $U_A(1)$ symmetry in the chiral restoration phase, we firstly analyze the susceptibility in chiral limit and at zero temperature. In this case, the disappeared chiral condensate $\langle\sigma\rangle_c=0$ leads to $\langle\sigma_\eta^2\rangle = \langle\sigma_{\eta'}^2\rangle =0$ and in turn
\begin{equation}
\chi_C^{(1)}=\chi_C^{(2)}=\chi_C^{(3)}=0,
\end{equation}
and the disappeared thermal excitation of mesons $n_B=0$ results in $J_a=I_{ab}=I_{abc}=0$ and in turn
\begin{equation}
\chi_C^{(2)}=\chi_C^{(3)}=\chi_C^{(4)}=\chi_M^{(1)}=\chi_M^{(2)}=0.
\end{equation}
This indicates that the strange condensate $\langle\sigma\rangle_s$ does not affect $U_A(1)$ symmetry, and the two broken symmetries, chiral symmetry and $U_A(1)$ symmetry, are simultaneously restored at the critical chemical potential $\mu_B^c$. In the real world with explicit chiral symmetry breaking at zero temperature, the nonzero chiral and strange condensates $\langle\sigma\rangle_c$ and $\langle\sigma\rangle_s$ lead to $\chi=\chi_C^{(1)}\ne 0$ in the chiral restoration phase. However, from our numerical calculation the value is very small, and $U_A(1)$ symmetry is almost restored, as shown in the panel with $T=0$ in Fig. \ref{fig4}.

Considering the significant mass splittings among the four independent mesons depicted in Fig. \ref{fig3} during the chiral restoration phase, the susceptibility calculation (\ref{wv}) via meson masses at the mean field level appears more problematic at finite density. It has been suggested that this approach should not be extended to finite temperature~\cite{Horvatic:2007qs,Benic:2011fv}.

As the temperature increases and the condensate part $\chi_C$ gradually melts, the meson fluctuation part $\chi_M$ is enhanced by the thermal motion. As a result of the competition, the total susceptibility decreases to the low temperature region, where the condensates are strong and thermal fluctuation is weak. Subsequently, it increases to the high temperature region, where the condensates become weak and thermal fluctuation becomes strong. In contrast to this non-monotonic temperature behavior, increasing density at any fixed temperature reduces the condensates and thermal fluctuation, the topological susceptibility is suddenly (at low temperature) or smoothly (at high temperature) suppressed by the density effect, and the $U_A(1)$ symmetry is restored at high baryon density. In the temperature and density evolution of the susceptibility, the role of the strange condensate is always weak, and the $U_A(1)$ restoration occurs at almost the same critical point as the chiral restoration. These features are clearly shown in Fig. \ref{fig4}.

\section{Summary and outlook}
\label{s4}
Considering the fact that chiral symmetry is broken at the classical level and $U_A(1)$ symmetry is broken at the quantum level, the mechanisms for the symmetry restorations are expected to differ. In this study, we investigated the relation between the two symmetries in the $SU(3)$ quark-meson model at finite temperature and baryon density. The topological susceptibility which describes the degree of the $U_A(1)$ breaking contains two components: the meson condensate controlled component and meson fluctuation component. As the temperature increases and condensates melt, the fluctuation becomes stronger. As a competition, the susceptibility behaves non-monotonically, and the $U_A(1)$ symmetry cannot be solely restored by the temperature effect. However, the density effect significantly differs. Specifically, it reduces both the condensates and fluctuation, and therefore the broken $U_A(1)$ symmetry can be restored only when the density effect is included. Although the strange condensate is still strong after the chiral phase transition and leads to large meson mass splittings at mean field level, its effect on the susceptibility, which is beyond the mean field, is very weak, and the two phase transitions, the chiral restoration and $U_A(1)$ restoration, occur at almost the same critical point. Based on the comparison of the Feynman diagrams for the susceptibility with the NJL model, the aforementioned qualitative conclusions appear to be independent of the model.

The $U_A(1)$ symmetry appears challenging to be restored in ultra-relativistic heavy ion collisions at the Large Hadron Collider (LHC), because the created medium is extremely hot but the baryon density can be neglected. However, in intermediate energy nuclear collisions and compact stars where the baryon density of the matter is high and the temperature is low, the $U_A(1)$ restoration can potentially be realized.

The calculations presented in this study involve certain approximations. Both quarks and mesons are approached using a mean field approximation, while the susceptibility is computed beyond the mean field, incorporating thermal fluctuations. This type of perturbative approach may introduce inconsistencies in the computation. Given that the meson propagators in the susceptibility remain at the mean field level, the baryon density effect manifests solely in the meson masses. Contributions from the quark-loop to the meson propagators are overlooked. Although thermal fluctuations are considered, the vacuum fluctuations, anticipated to be significant in a dense medium, are omitted from the analysis.

\section{Acknowledgments}
We are grateful to Profs. Lianyi He and Yin Jiang for helpful discussions during our study. The work is supported by the Grants NSFC11890712, NSFC12075129, 2018YFA0306503 and 2020B0301030008.

\bibliographystyle{apsrev4-2}
\bibliography{Ref}	
\end{document}